\documentclass[aps,prb,twocolumn,amsmath,amssymb,floatfix,superscriptaddress]{revtex4-1}
\usepackage{tabularx}
\usepackage{bm}
\usepackage{euscript}
\usepackage{epsfig,psfrag,subfigure}
\usepackage{graphicx}
\usepackage{color}
\usepackage{amsfonts}
\usepackage{exscale}
\usepackage{amsbsy}
\usepackage{multirow}
\usepackage{hyperref}
\usepackage[dvipsnames]{xcolor}
\hypersetup{colorlinks=false,linkcolor=black}
\usepackage[]{lineno}
\usepackage{comment}

\begin{document}
\title{Doping quantum spin liquids on the Kagome lattice}

\author{Cheng Peng}
\thanks{These authors contributed equally.}
\affiliation{Stanford Institute for Materials and Energy Sciences, SLAC National Accelerator Laboratory and Stanford University, Menlo Park, California 94025, USA}

\author{Yi-Fan Jiang}
\thanks{These authors contributed equally.}
\affiliation{Stanford Institute for Materials and Energy Sciences, SLAC National Accelerator Laboratory and Stanford University, Menlo Park, California 94025, USA}

\author{Dong-Ning Sheng}
\affiliation{Department of Physics and Astronomy, California State University, Northridge, California 91330, USA}

\author{Hong-Chen Jiang}
\email{hcjiang@stanford.edu}
\affiliation{Stanford Institute for Materials and Energy Sciences, SLAC National Accelerator Laboratory and Stanford University, Menlo Park, California 94025, USA}

\begin{abstract}
We review recent density-matrix renormalization group (DMRG) studies of lightly doped quantum spin liquids (QSLs) on the kagome lattice. While a number of distinct conducting phases, including high temperature superconductivity, have been theoretically anticipated we find instead a tendency toward fractionalized insulating charge-density-wave (CDW) states. In agreement with earlier work (Jiang, Devereaux, and Kivelson, Phys. Rev. Lett. ${\bf 119}$, 067002 (2017)),   results for the $t$-$J$ model reveal that, starting from a fully gapped QSL, light doping leads to CDW long-range order with a pattern that depends on lattice geometry and doping concentration such that  there is one doped-hole per CDW unit cell, while the spin-spin correlations remain short-ranged.  Alternatively, this state can be viewed as a stripe crystal or Wigner crystal of spinless holons, rather than of doped holes. From here, by studying generalized versions of the $t$-$J$ model, we extend these results to light doping of other types of QSLs, including critical and chiral QSLs. Our results suggest that doping these QSLs also leads to insulating states with long-range CDW order. While the superconducting correlations are short-ranged, they can be significantly enhanced by second-neighbor electron hopping. The relevance of our numerical results to Kagome materials is also discussed.
\end{abstract}

\maketitle

\section{Introduction}
Quantum spin liquids (QSLs) are highly entangled phases of matter that exhibit a variety of novel features associated with their topological character, resisting symmetry breaking even at zero temperature due to strong quantum fluctuations and geometric frustrations.\cite{Anderson1973,Moessner2001,Balents2010,Savary2016,Broholm2019} Theoretically, there are distinct types of QSLs which correspond to different patterns of long-range entanglement, and can be distinguished based on whether their excitation spectrum above the ground state has a gap or not, i.e., gapped or gapless QSLs.\cite{Broholm2019} Broad interest in QSLs was triggered by the notion that they can be viewed as insulating phases with preexisting electron pairs, such that upon light doping they might automatically yield high temperature superconductivity.\cite{Anderson1987,Kivelson1987,Rokhsar1988,Laughlin1988,Wen1989,Wen1996,Broholm2019} It has also been proposed that doped QSLs might form distinct types of conducting phases, including fractional Fermi liquid. \cite{Senthil2003,Punk2015,Lee2006,Fradkin2015,Broholm2019} While these illustrate the importance of the spin liquids as the parent states, it is also worth noticing that  the realization of QSL is a great challenge to physicists and candidate materials are rare.\cite{Lee2006,Balents2010,Broholm2019}

The spin-1/2 antiferromagnetic (AFM) Heisenberg model on the kagome lattice with nearest-neighbor (NN) superexchange 
interaction is geometrically frustrated. A number of numerical simulations\cite{Jiang2008,Yan2011,Jiang2012,Depenbrock2012,Gong2015,Mei2017,Ran2007,Clark2013,Iqbal2013,Liao2017,He2017} have provided strong evidences that its ground state is a QSL. Experimentally, the celebrated materials herbertsmithite ZnCu$_3$(OH)$_6$Cl$_2$ and Zn-substituted barlowite Cu$_3$Zn(OH)$_6$FBr are promising realizations of the kagome antiferromagnet.\cite{Han2012,Fu2015,Liu2015,Hering2017,Feng2017,Smaha2020} Experimental evidence of fractional spin excitations has been observed in neutron scattering and strong indications of a spin gap are seen in NMR studies of single crystals.\cite{Han2012,Fu2015}
By introducing both second- and third-neighbor Heisenberg interactions\cite{Messio2012}, recent density-matrix renormalization group (DMRG) studies have identified a distinct chiral spin liquid (CSL), which spontaneously breaks time-reversal symmetry (TRS)\cite{Gong2014,He2014,Wen1989,Yang1993,Haldane1995,Laughlin1983,Kalmeyer1987}. Moreover, this CSL can also be realized in the anisotropic XXZ spin model involving both second- and third-neighbor coupling,\cite{He2015,Zhu2015} or by explicitly introducing the TRS breaking three-spin scalar chiral interaction.\cite{Bauer2014,Hu2015}

This raises a natural question of whether doping QSLs yield superconductivity. Recent DMRG studies have provided strong evidences that lightly doping the time-reversal-symmetric QSL on the triangular lattice can naturally yield nematic $d$-wave superconductivity.\cite{Jiang2019} Similarly, more DMRG studies have also shown that the topological $d\pm id$-wave superconductivity can be realized in doped CSL on the triangular lattice.\cite{Jiang2020CSL} While high temperature superconductivity has been theoretically anticipated and numerically realized in doping QSLs on the triangular lattice, we find instead a tendency toward fractionalized insulating charge-density-wave (CDW) states in doping QSL on the kagome lattice. It has been shown in a recent study in the context of $t$-$J$ model\cite{Jiang2017} that the doped holes form an insulating state with long-range CDW order, with patterns of unidirectional stripe crystal or two-dimensional (2D) Wigner crystal dependent on lattice geometry and doping concentrations. Specifically, this CDW state appears exotic with spin-charge separation which is formed by spinless holons instead of the original doped holes.

In this paper, by studying generalized versions of the $t$-$J$ model, we extend these results to light doping of other types of QSLs, including critical and chiral QSLs. In agreement with our earlier work,\cite{Jiang2017} our results suggest that doping these QSLs also leads to insulating states with long-range CDW order, which is formed by fractionalized spinless holons instead of original doped holes. This is evidenced by that fact that the doped system can be approximately divided into emergent larger CDW unit cells with one doped hole, while the spin-spin correlations remain short-ranged and there is no spin accumulation around each doped hole. Table \ref{Table:Summary} summarizes the main results obtained by DMRG and the details will be discussed in the following.

This paper is organized as follows. In Section II, we review the previous study of lightly doped QSL of the AFM Heisenberg model on the kagome in the context of the $t$-$J$ model and show more results by introducing additional second-neighbor electron hopping terms. In Section III, we present a new study of lightly doped critical spin liquid of the AFM XY model on the kagome lattice. In Section IV, we present a study of lightly doped chiral spin liquid on the Kagome lattice. Section V is devoted to summary and conclusion.

\begin{figure}
  \includegraphics[width=\linewidth]{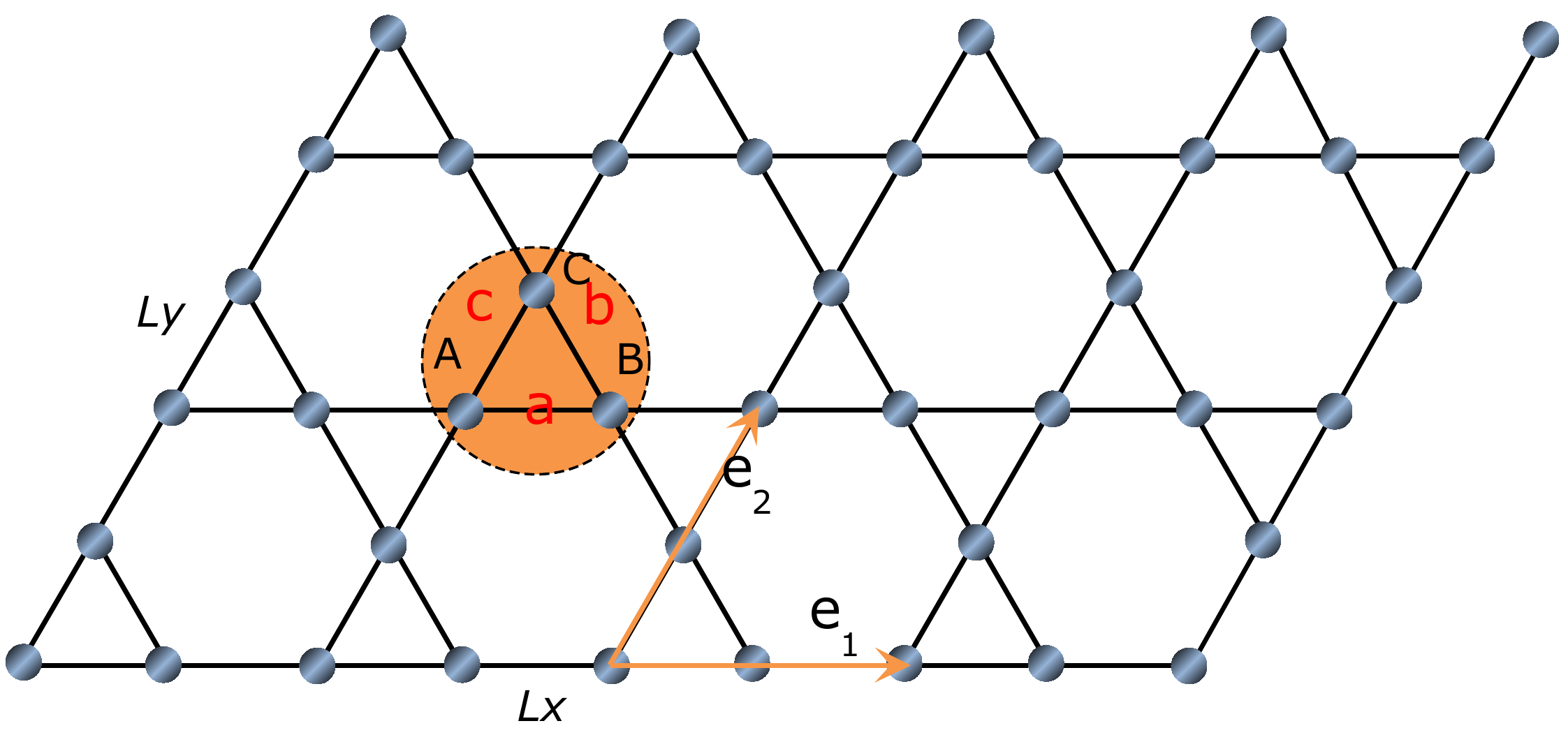}
  \caption{(Color online) The geometry of Kagome cylinder used in Ref.\cite{Jiang2017} and this paper. Open and periodic boundary conditions are imposed along the basis vectors $\bold{e}_1$ and $\bold{e}_2$ directions, respectively. Each unit cell (denoted by small triangle in the shaded region) has three sites (A, B and C) and three bonds (a, b and c). The cylinder shown in this figure (refereed to as $\rm YC$--$2L_y$--$L_x$) has $L_x$ and $L_y$ unit cells in the $\bold{e}_1$ and $\bold{e}_2$ directions, respectively. Note that the unit cells at the right boundary are incomplete in order to reduce boundary effects in the DMRG simulation.}\label{Fig:Lattice}
\end{figure}

\begin{table*}[tb]
\centering 
\begin{tabular}{c | c | c | c | c | c | c | c | c } 
\hline\hline 
QSL & Model  &  Parameters & Kagome cylinder & Doping region & Phase &  $A_{cdw}$ & $\xi_s$ & $\xi_{sc}$ \\ [.5ex] 
\hline 
 Gapped  & $t$-$J$\cite{Jiang2017} & $t$=3, $J$=1  & YC-6, YC-8 & $\delta \leq 1/15$ & Stripe/Wigner crystal & 0.009 -- 0.025 & 1.1 -- 2.1 & 0.5 -- 1.3 \\
 Gapped  &  $t$-$t^\prime$-$J$ & $t$=3, $J$=1, $t^\prime$=0--2 & YC-6 & $\delta=1/18$ & Stripe crystal &   &   & 0.7 -- 3.0   \\
 Critical& $t$-$J_{xy}$ & $t$=3, $J_{xy}$=1 & YC-6, YC-8 & $\delta\leq 1/18$ & Stripe/Wigner crystal & 0.006 -- 0.021 & 1.4 -- 2.1 & 0.6 -- 0.8 \\
 Chiral & $t$-$J$-$J_\chi$ & $t$=3, $J$=1, $J_{\chi}$=0.5 & YC-6, YC-8 & $\delta\leq 1/12$ & Stripe crystal & 0.018 -- 0.025  & 1.1 -- 1.6  &   0.8 -- 1.1 \\
\hline\hline 
\end{tabular}
\caption{Summary of the phases. QSL denotes the type of parent spin liquids before doping. Corresponding model, parameters, doping region $\delta$, phase characters, CDW order parameter $A_{cdw}$, spin-spin $\xi_s$ and SC correlation lengths $\xi_{sc}$ in the unit of lattice spacing. Stripe and Wigner crystal means the phase with long-range unidirectional and 2D charge-density-wave order, respectively.}\label{Table:Summary}
\end{table*}

\section{Doping gapped spin liquid}\label{DGSL} %
The physics of lightly doped QSL of the AFM Heisenberg model on the kagome lattice has been numerically addressed by Jiang, Devereaux and Kivelson in a recent study.\cite{Jiang2017} Large-scale DMRG simulations show that doping such a spin liquid leads to an insulating state with long-range CDW order, with patterns of unidirectional stripe crystal or 2D Wigner crystal dependent on lattice geometry and doping concentration. Interestingly, the crystal is formed of spinless holons rather than doped holes, which only carry charge degree freedom of electron. Approximately, the doped system can be divided into emergent larger unit cells with one doped hole, however, the spin-spin correlations decay exponentially without spin accumulation around each hole. This provides an explicit example of spin-charge separation realized in systems beyond one dimension.

Specifically, the ground state properties of lightly doped QSL on the kagome lattice has been calculated in the context of $t$-$J$ model using DMRG,\cite{White1992} which is defined by the Hamiltonian%
\begin{eqnarray}
H = -t\sum_{\langle ij\rangle\sigma}(\hat{c}^+_{i\sigma}\hat{c}_{j\sigma}+h.c.) +J\sum_{\langle ij\rangle}(\mathbf{S}_i\cdot \mathbf{S}_j-\frac{\hat{n}_i\hat{n}_j}{4}). \label{Eq:HamtJ}
\end{eqnarray}
Here $\hat{c}^+_{i\sigma}$($\hat{c}_{i\sigma}$) is the electron creation (annihilation) operator with spin-1/2 on site $i$, $\mathbf{S}_i$ is the spin operator and $\hat{n}_i=\sum_\sigma \hat{c}^+_{i\sigma}\hat{c}_{i\sigma}$ is the electron number operator. $\langle ij\rangle$ denotes NN sites and the Hilbert space is constrained by the no-double occupancy conditions $n_i\leq 1$. At half-filling, i.e., $n_i=1$, the $t$-$J$ model reduces to the spin-1/2 AFM Heisenberg model whose ground state is a QSL. Although it is still under debate whether the spin liquid is gapped or gapless in 2D\cite{Jiang2008,Yan2011,Jiang2012,Depenbrock2012,Gong2015,Mei2017,Ran2007,Clark2013,Iqbal2013,Iqbal2014,Liao2017,He2017}, the fact that the observed spin correlation lengths are short compared to the width of the cylinders is consistent with the properties of a spin-gapped state on finite-width cylinders.

\begin{figure}
  \includegraphics[width=\linewidth]{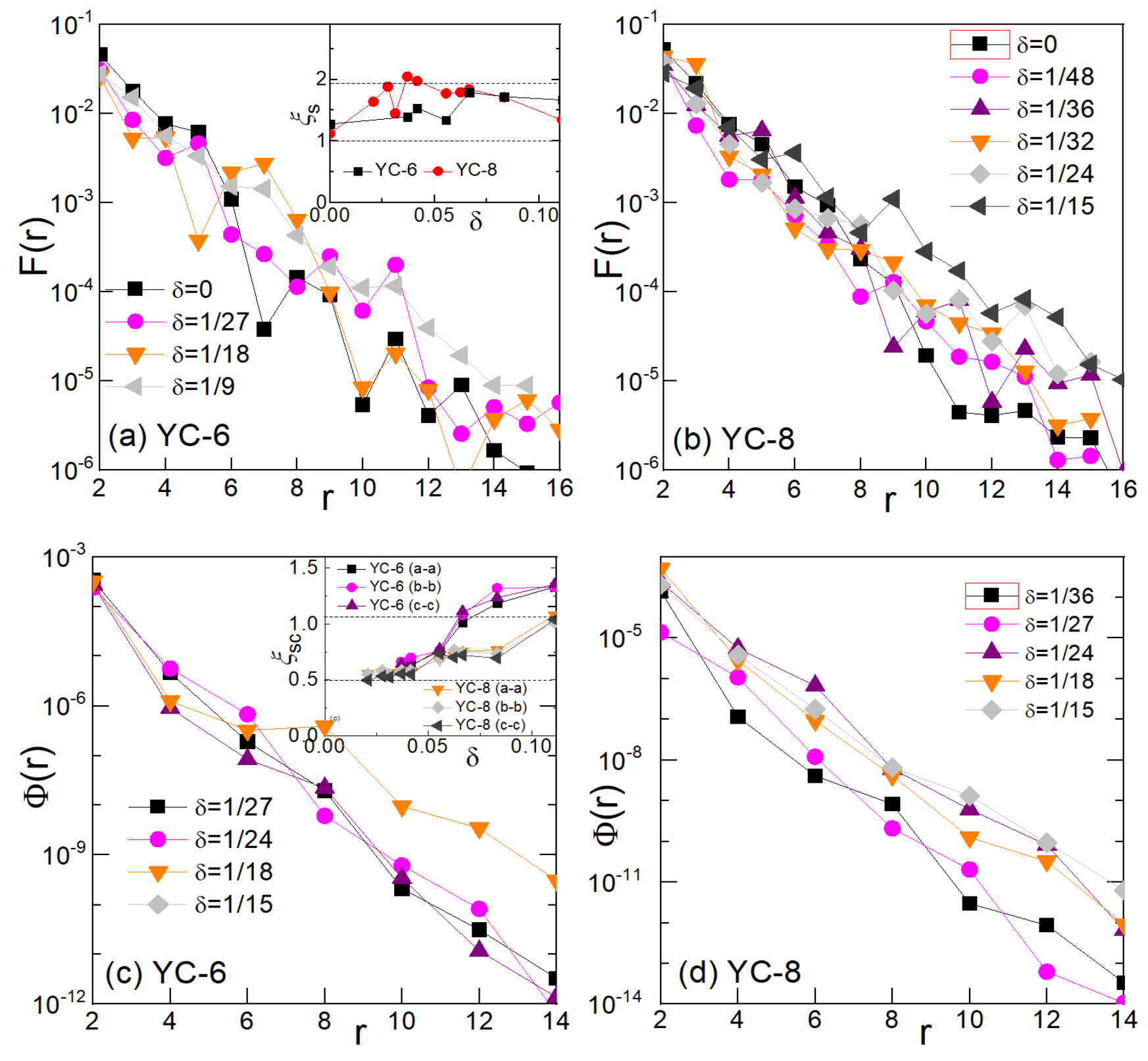}
  \caption{(Color online) The spin-spin $F(r)$ (a-b) and superconducting $\Phi(r)$ (c-d) correlation functions for $\rm YC$-6 and $\rm YC$-8 cylinders at different hole doping concentrations $\delta$. $r$ is the distance between two sites in the $\bold{e}_1$ direction. Insets: The spin-spin $\xi_s$ and superconducting $\xi_{sc}$ correlation lengths as a function of $\delta$. Results are from Ref.\cite{Jiang2017}.}  \label{Fig:tJ_Spin_SC_Cor}
\end{figure}

The DMRG simulations were performed on YC--2$L_y$--$L_x$ cylinders shown in Fig.\ref{Fig:Lattice}, where $L_y$ and $L_x$ are the numbers of unit cells ($2L_y$ and $2L_x+1$ are the number of sites) along the $\bold{e}_2$ and $\bold{e}_1$ directions, respectively. Note that one more column at the right boundary with only two sites (A and C) are added in order to reduce boundary effect due to sharp edges. The total number of sites is $N=L_y(3L_x+2)=N_u+2L_y$, where $N_u$ is the number of sites inside intact unit cells. The doping concentration of the system away from half-filling is defined as $\delta = N_h/N_u$, where $N_h$ is the number of doped holes. Although $N_u\neq N$ so that the average value of $\delta$ differs slightly from $\tilde{\delta}=N_h/N$, deep in the bulk, i.e., if relatively far from the open boundaries there has $\tilde{\delta}=\delta$.

One of the main results is that the spin-spin correlation functions, defined as%
\begin{eqnarray}
F(r)=\frac{1}{L_y} \sum_{y=1}^{L_y} |\langle \bold{S}_{i_0} \cdot \bold{S}_{i_0+r}\rangle|, \label{Eq:Spin_Cor}
\end{eqnarray}
are remarkably insensitive to doping concentration $\delta$. Here $\bold{S}_{i_0}$ is the spin operator on the reference site $i_0=(x_0,y)$ in the middle of the cylinders and $r$ is the distance between two sites in the $\bold{e}_1$  direction. This is very similar with the half-filled case $\delta=0$ for both YC-6 and YC-8 cylinders, where the ground state of the system is a QSL with short-range spin-spin correlations.\cite{Jiang2008,Yan2011,Jiang2012,Depenbrock2012,Kolley2015} For both cases, $F(r)$ decays exponentially at long distances and can be well fitted by an exponential function $F(r)\sim e^{-r/\xi_{s}}$ with short correlation lengths $\xi_s$=1$\sim 2$ lattice spacings as shown in Fig.\ref{Fig:tJ_Spin_SC_Cor}(a-b).\cite{Jiang2017}

\begin{figure}
  \includegraphics[width=\linewidth]{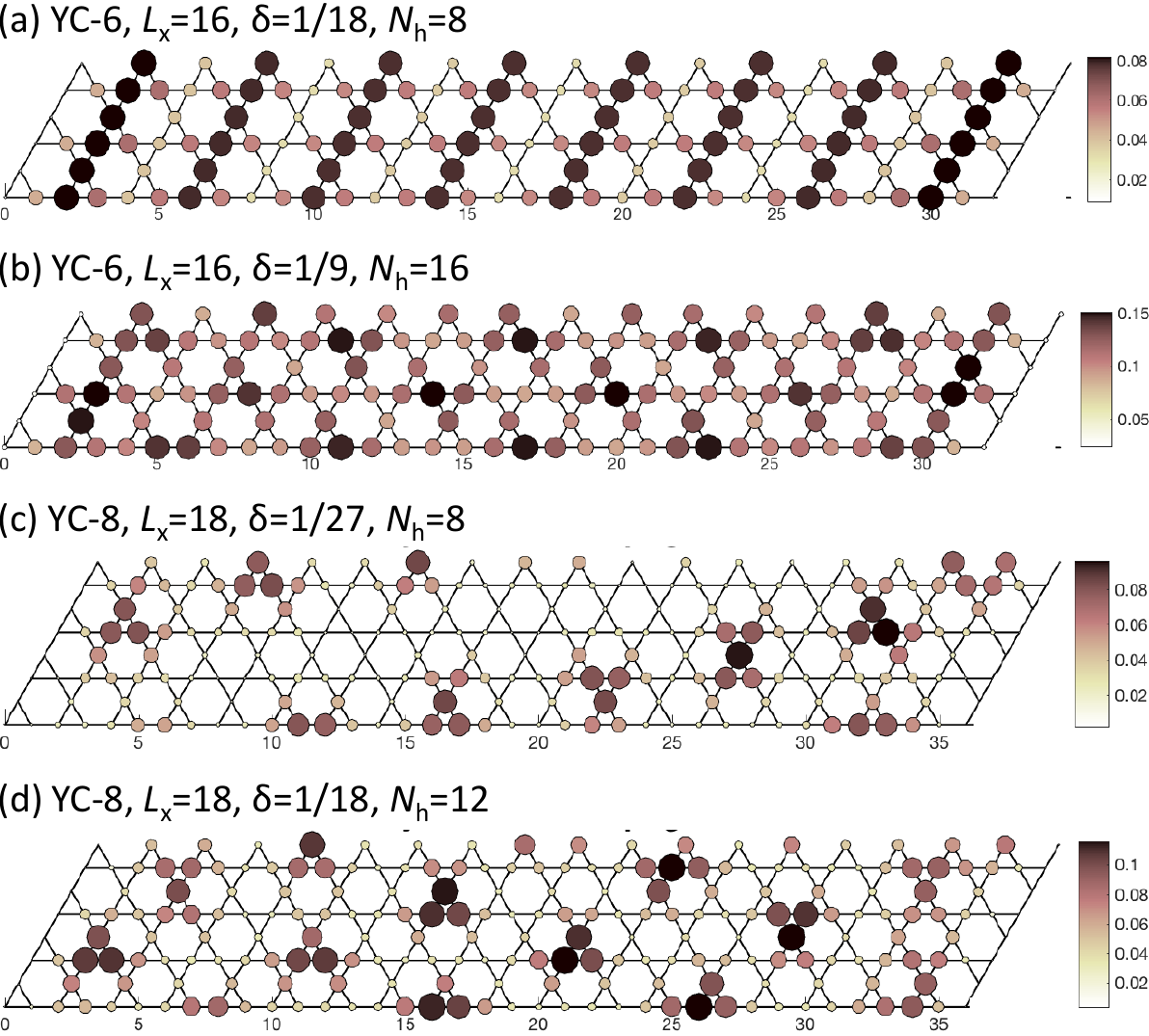}
  \caption{(Color online) The charge density profile $n_h(x,y)$ at different hole doping concentrations $\delta$ for $\rm YC$-6 in (a-b) and $\rm YC$-8 cylinders in (c-d). Results are from Ref.\cite{Jiang2017}.}\label{Fig:tJ_Ni}
\end{figure}

Another main conclusion is that the superconducting (SC) correlations, defined as%
\begin{eqnarray}
\Phi_{\alpha \beta}(r)=\frac{1}{L_y}\sum_{y=1}^{L_y}|\langle \Delta^\dagger_\alpha(i_0) \Delta_\beta(i_0+r)\rangle|,\label{Eq:SC_Cor}
\end{eqnarray}
are also short-ranged, which is similar with the spin-spin correlations $F(r)$. Here, $\Delta^\dagger_\alpha(i)$ is the spin-singlet pair-field creation operator given by
$\Delta_\alpha^\dagger(i) = \frac{1}{\sqrt{2}} \left( \hat{c}^\dagger_{i \uparrow} \hat{c}^\dagger_{i + \alpha \downarrow} - \hat{c}^\dagger_{i \downarrow} \hat{c}^\dagger_{i + \alpha \uparrow} \right)$,
where $\alpha$=a/b/c denotes the bond type as shown in Fig.\ref{Fig:Lattice}) with bond vectors defined as $\bold{a}=\bold{e}_1/2$, $\bold{c}=\bold{e}_2/2$ and $\bold{b}=(\bold{e}_2 - \bold{e}_1)/2$. $i_0$ is the index of the reference bond, and $r$ is the distance between two bonds along the $\bold{e}_1$ direction. Specifically, all the SC correlations $\Phi_{\alpha\beta}(r)$ decay exponentially at long distances for both YC-6 and YC-8 cylinders and can be well fitted by the exponential function $\Phi(r)\sim e^{-r/\xi_{sc}}$, as shown in Fig.\ref{Fig:tJ_Spin_SC_Cor}(c-d). The SC correlation lengths $\xi_{sc}=0.5\sim 1.3$ are small for all doping concentrations $0< \delta \leq 11\%$ that has been explored in Ref.\cite{Jiang2017}. This is qualitatively distinct with the theoretical study for the presence of topological superconductivity in the lightly-doped kagome QSL.\cite{Ko2009}

Contrary to both spin-spin and SC correlations, CDW correlations appear to be long-ranged in the lightly doped QSL on the Kagome lattice. Examples of the charge density profile $n_h(x,y)=1-n(x,y)$ are shown in Fig.\ref{Fig:tJ_Ni}, where $n(x,y)$ is the electron density on site $i=(x,y)$. A clear CDW ordering is observed, although its pattern, i.e., unidirectional stripe or 2D crystal, depends on either the lattice geometry and doping concentration. However, for both cases, the doped system can be divided into new larger emergent unit cells, each has one of the red stripes in Fig.\ref{Fig:tJ_Ni}(a) or one of the red-spots in Fig.\ref{Fig:tJ_Ni}(b-d). Specifically, the number of emergent unit cells is equal to the number of doped holes at all doping concentrations, showing that there are no Cooper pairs. It is worth mentioning that the CDW order appears to be long-ranged ordered, whose existence can be determined by fitting the amplitude $A_{cdw}$ of the rung charge density $n_h(x)=\frac{1}{L_y}\sum_{y=1}^{L_y}n_h(x,y)=A_{cdw} \cos (Q x+\theta)+B$, where $Q$ and $B$ are fitting parameters. As shown in Ref.\cite{Jiang2017}, $A_{cdw}$ remains finite in the $L_x=\infty$ limit.\cite{Jiang2017}. This is distinct with the quasi-long-range CDW order in the lightly doped QSLs on the triangular cylinders\cite{Jiang2019,Jiang2020CSL}.

\begin{figure}
  \includegraphics[width=\linewidth]{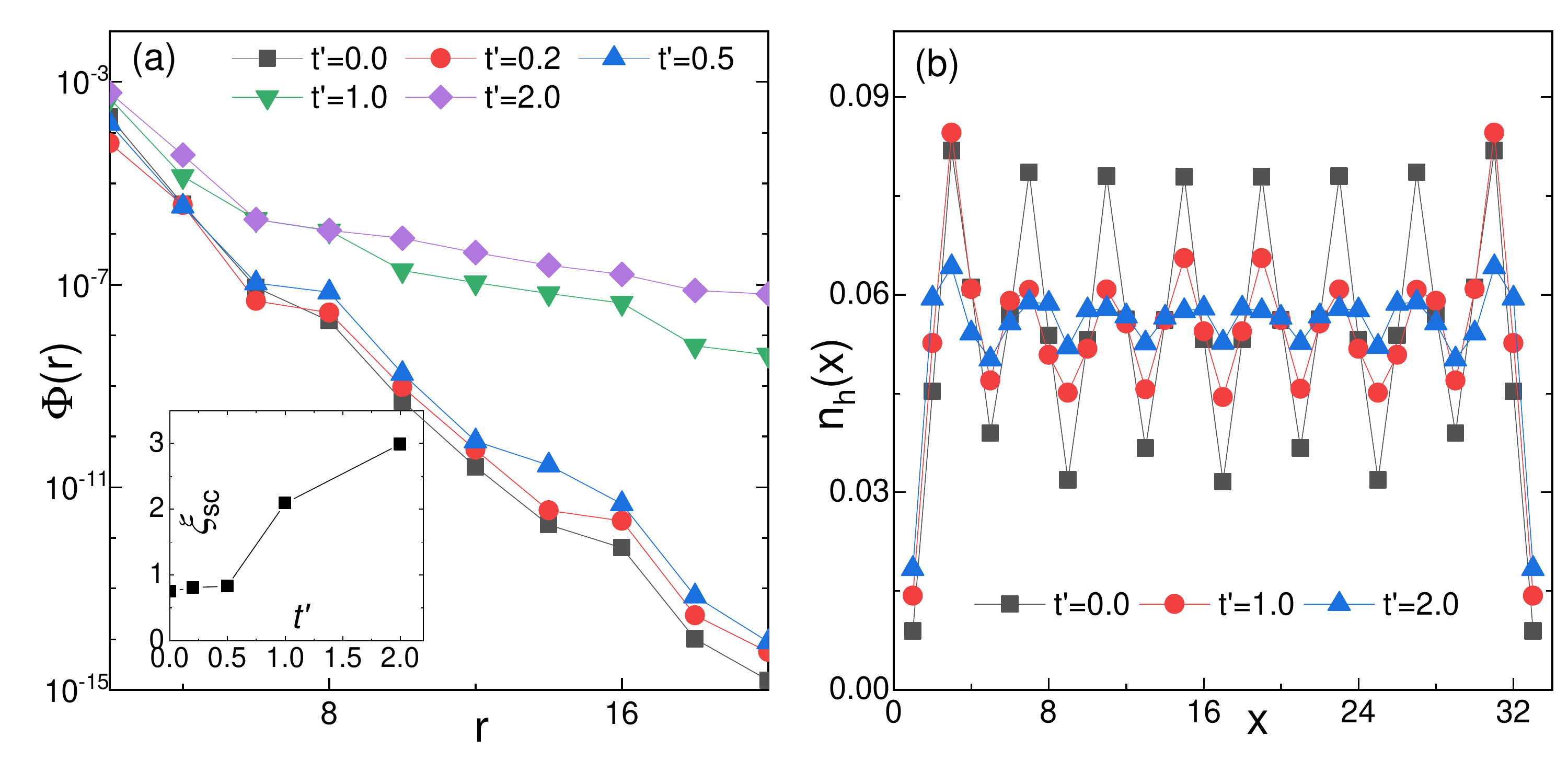}
  \caption{(Color online) (a) Superconducting correlations and (b) charge density profile $n_h(x)$ for YC-6 cylinders as a function of second-neighbor electron hopping $t^\prime$ at $\delta=1/18$ hole doping concentration. Inset: Extracted superconducting correlation length $\xi_{sc}$ as a function of $t^\prime$.} \label{Fig:tJ_SC_Ni_tp}
\end{figure}

\textit{Effect of second-neighbor hopping $t^\prime$:} %
It has been shown in previous studies\cite{Jiang2019Hub,Jiang2020} that the second-neighbor electron hopping $t^\prime$ of the Hubbard model on the square lattice is important to enhance superconductivity and suppress CDW order. In this paper, we will show that this is also true for the lightly doped QSL on the Kagome lattice. To demonstrate this, we have studied the $t$-$J$ model in Eq.(\ref{Eq:HamtJ}) with additional second-neighbor electron hopping term $-t^\prime \sum_{\langle\langle ij\rangle\rangle\sigma}(c^+_{i\sigma}c_{j\sigma}+h.c.)$. In our DMRG simulations, we have kept up to $m=18000$ number of states in each DMRG block with a typical truncation error $\epsilon\sim 10^{-6}$. Examples for YC-6 Kagome cylinders at $\delta=1/18$ hole doping concentration are given in Fig.\ref{Fig:tJ_SC_Ni_tp}. As expected, it is clear that the presence of $t^\prime$ can significantly enhance the SC correlations $\Phi(r)$ as shown in Fig.\ref{Fig:tJ_SC_Ni_tp}(a). For instance, the value of $\Phi(r)$ can be enhanced by many orders of magnitudes when $t^\prime$ is comparable with $J$ and even more for larger $t^\prime$. While the SC correlations $\Phi(r)$ still decay exponentially at long distances which can be fitted by exponential function $\Phi(r)\sim e^{-r/\xi_{sc}}$, the SC correlation length becomes significantly longer with the increase of $t^\prime$, as shown in the inset of Fig.\ref{Fig:tJ_SC_Ni_tp}(a). On the contrary, the CDW order gets significantly suppressed as evidenced by the charge density profle $n_h(x)$ for the same systems as shown in Fig.\ref{Fig:tJ_SC_Ni_tp}(b). The oscillation amplitude $A_{cdw}$ of $n_h(x)$ becomes significantly smaller as $t^\prime$ increases. Similar with previous studies of Hubbard model on the square lattice,\cite{Jiang2019Hub,Jiang2020} our results demonstrate the importance of second-neighbor electron hopping $t^\prime$ which provides a promising pathway to potentially realize superconductivity in doping QSL on the Kagome lattice. However, it is also worth mentioning that even the superconducting correlations can be significantly enhanced by $t^\prime$, quasi-long-range superconductivity has not yet been realized on the Kagome lattice. This is qualitatively distinct with the square lattice, which may suggest that even longer range hopping terms are required to realize superconductivity on the Kagome lattice.

\begin{figure}
\centering
    \includegraphics[width=1\linewidth]{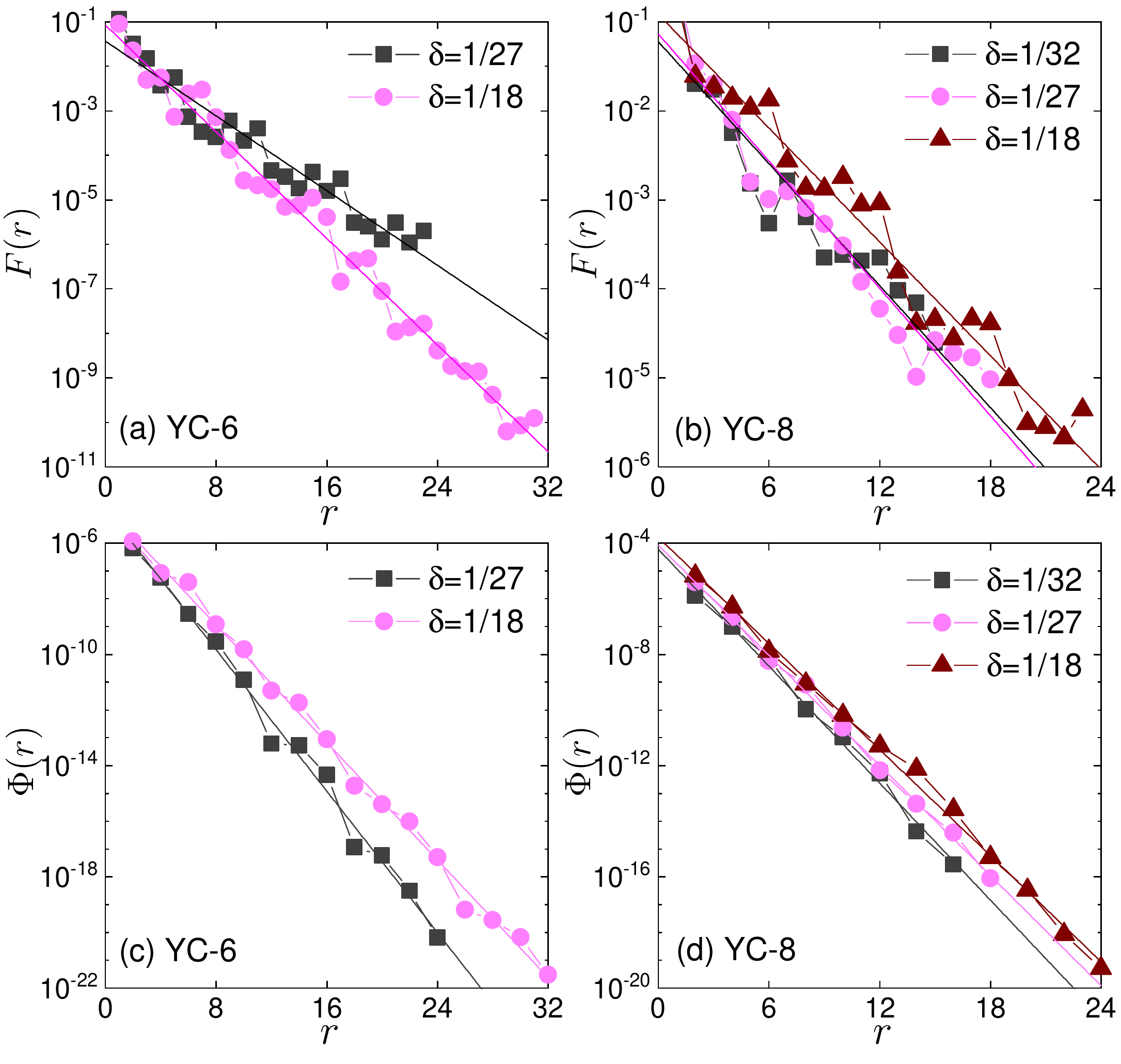}
\caption{(Color online) The spin-spin correlation functions $F(r)$ along the $e_1$ direction for (a) YC-6 and (b) YC-8 cylinders, and the superconducting pair-field correlation functions $\Phi(r)$ along the $e_1$ direction between a-a bonds for (c) YC-6 and (d) YC-8 cylinders at different hole doping concentrations $\delta$.} \label{Fig:DCriticalSL_Corl}
\end{figure}

\section{Doping critical spin liquid}\label{DCriSP}
Apart from the gapped spin liquid, critical spin liquid with gapless spin singlet excitation is another type of QSL that can be realized on the Kagome lattice. \cite{Ran2007,Iqbal2013,Iqbal2014,Hastings2000,Rantner2001} A number of numerical simulations have established that the AFM XY model on the Kagome lattice has the critical spin liquid ground state.\cite{Zhu2015,Hu2015} While it has been theoretically proposed that doping gapless spin liquid could yield superconductivity,\cite{Anderson1973,Anderson1987,Hermele2005} direct  evidences based on unbiased numerical study remains still lacking. Meanwhile, it is also interesting to explore whether other nontrivial or metallic states of matter could emerge in doping such a gapless spin liquid. To address this question, we employ the DMRG to investigate the ground state properties of the lightly hole-doped extended $t$-$J$ model on the Kagome lattice, i.e., $t$-$J_{xy}$ model, where the Hamiltonian is defined as%
\begin{equation}
H= -t\sum_{\langle ij\rangle \sigma}(\hat{c}^{\dagger}_{i\sigma}\hat{c}_{j\sigma}+h.c.) + \frac{J_{xy}}{2}\sum_{\langle ij\rangle}(S^{+}_i \cdot S^{-}_j+h.c.) \label{Eq:HamtJxy}
\end{equation}
Here $\hat{c}^{\dagger}_{i\sigma}$($\hat{c}_{i\sigma}$) is the electron creation(annihilation) operator with spin-$\sigma$ on site $i=(x_i,y_i)$, $S^{+}_i$($S^{-}_i$) is the creation(annihilation) of the $S=1/2$ spin operator on site $i$. $J_{xy}$ is the NN superexchange spin interaction. At half-filling, i.e., $n_i=1$, this $t$-$J_{xy}$ model reduces to the spin-1/2 AFM XY model, which has the critical spin liquid ground state.\cite{Zhu2015,Hu2015} For the present study, we mainly focus on the lightly doped case with hole doping concentration $0 < \delta\leq 1/18$. We set $J_{xy}=1$ as an energy unit and consider $t=3$. We keep up to $m=7000$ number of states in each DMRG block and perform up to $100$ sweeps with a typical truncation error $\epsilon \sim 10^{-6}$ for YC-$6$ cylinders, $\epsilon \sim 10^{-5}$ for YC-$8$ and YC-$10$ cylinders, respectively. This gives excellent convergence for the results after extrapolated to the  $\epsilon\rightarrow 0$ limit, i.e., $m\rightarrow\infty$.

\textit{Spin-spin correlations:} %
We calculate the spin-spin correlation functions $F(r)$ defined in Eq.(\ref{Eq:Spin_Cor}) to study the magnetic properties of the ground state. To minimize the boundary effect, we choose the reference site at $i_0=(x_0,y)$ with $x_0\sim L_x/4$ and calculate $F(r)$ with $r$ the distance between two sites along the $\bold{e}_1$ direction including both A and B sites. Upon light doping, we find the spin-spin correlation $F(r)$ becomes short-ranged for all the doping concentrations that we have considered in the present study, including $\delta=1/27$ and $1/18$ as shown in Fig.\ref{Fig:DCriticalSL_Corl}(a) for YC-6 and $\delta=1/32, 1/27$ and $1/18$ as shown in Fig.\ref{Fig:DCriticalSL_Corl}(b) for YC-8 cylinders, respectively. For all these cases, we find that $F(r)$ decays exponentially at long distances and can be well fitted by an exponentially falling function $F(r)\sim e^{-r/\xi_s}$ where $\xi_s$ is the spin-spin correlation length. Although $\xi_{s}$ slightly depends on $\delta$ and lattice geometry, it remains finite and small, e.g., $\xi_{s}=1.45(3)$--$2.1(2)$ lattice spacings for YC-$6$ and YC-$8$ cylinders. Therefore, we can conclude that there is no (quasi-) long-range magnetic order in the lightly doped critical spin liquid on the Kagome lattice. 

\begin{figure}
\centering
    \includegraphics[width=1\linewidth]{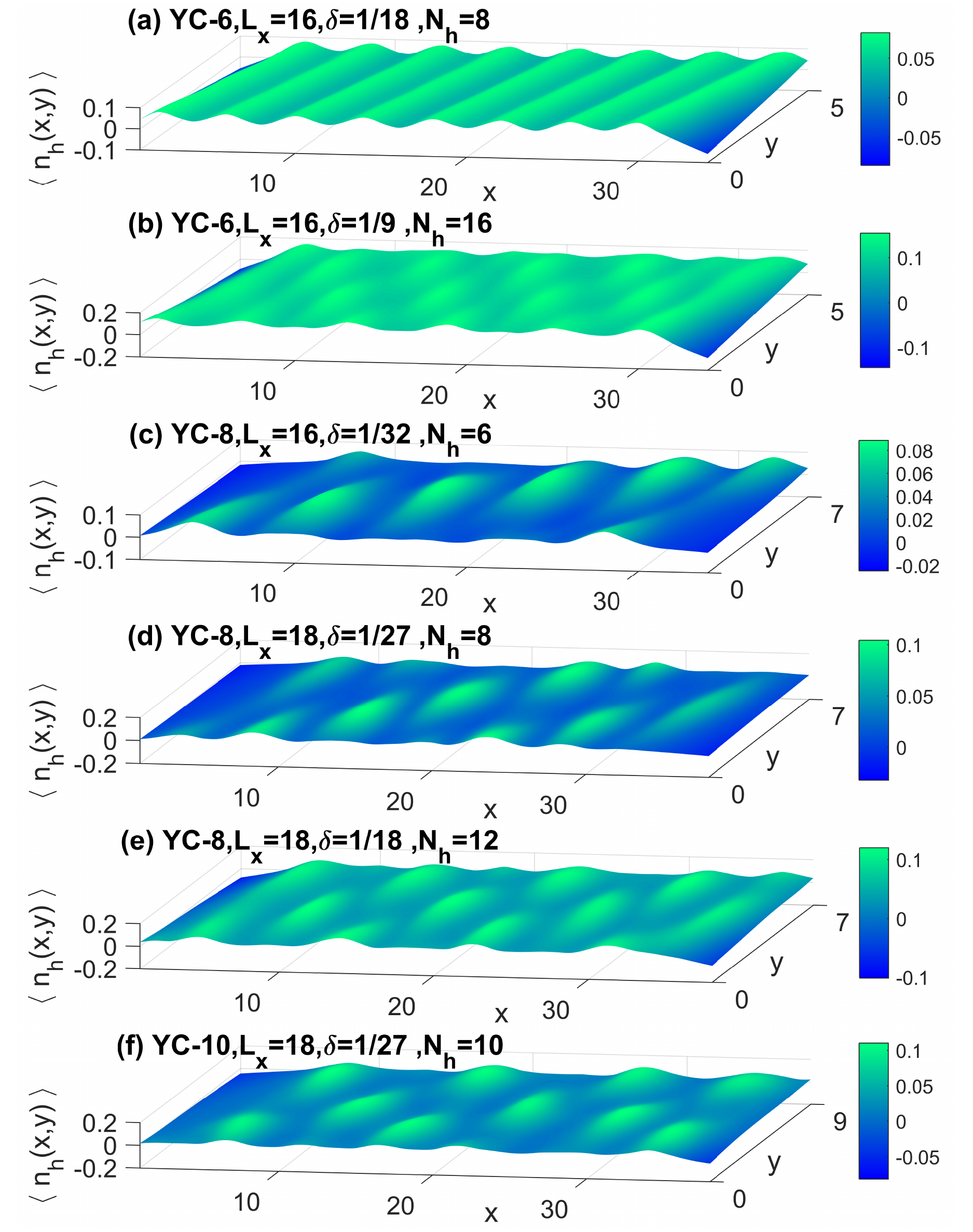}
\caption{(Color online) The charge density profile $n_h(x,y)$ at different hole doping concentrations $\delta$ for YC-$6$ in (a)-(b), YC-$8$ in (c)-(e) and YC-$10$ in (f).}\label{Fig:DCriticalSL_CDW}
\end{figure}

\textit{Superconducting correlations:} %
To test the possibility of superconductivity, we
have calculated the equal-time SC pair-field correlation functions defined in Eq.(\ref{Eq:SC_Cor}). As the ground state of the system with an even number of doped holes is always found to have zero spin, we will focus on spin-singlet pairing. Upon light doping the critical spin liquid, we find that all types of the SC correlations $\Phi_{\alpha\beta}(r)$ with $\alpha(\beta)=a/b/c$, decay exponentially at long distances for both YC-$6$ and YC-$8$ cylinders. As shown in Fig.\ref{Fig:DCriticalSL_Corl}(c-d), the SC pair-pair correlations $\Phi_{aa}(r)$ for three different doping concentrations can be well fitted by an exponential function $\Phi(r)\sim e^{-r/\xi_{sc}}$ with short SC correlation length $\xi_{sc}=0.62(2)$--$0.82(1)$ lattice spacings. Therefore, our results suggest that there is no (quasi-) long-range superconductivity in lightly doped critical spin liquid phase on the Kagome lattice. 

\begin{figure}
\centering
    \includegraphics[width=1\linewidth]{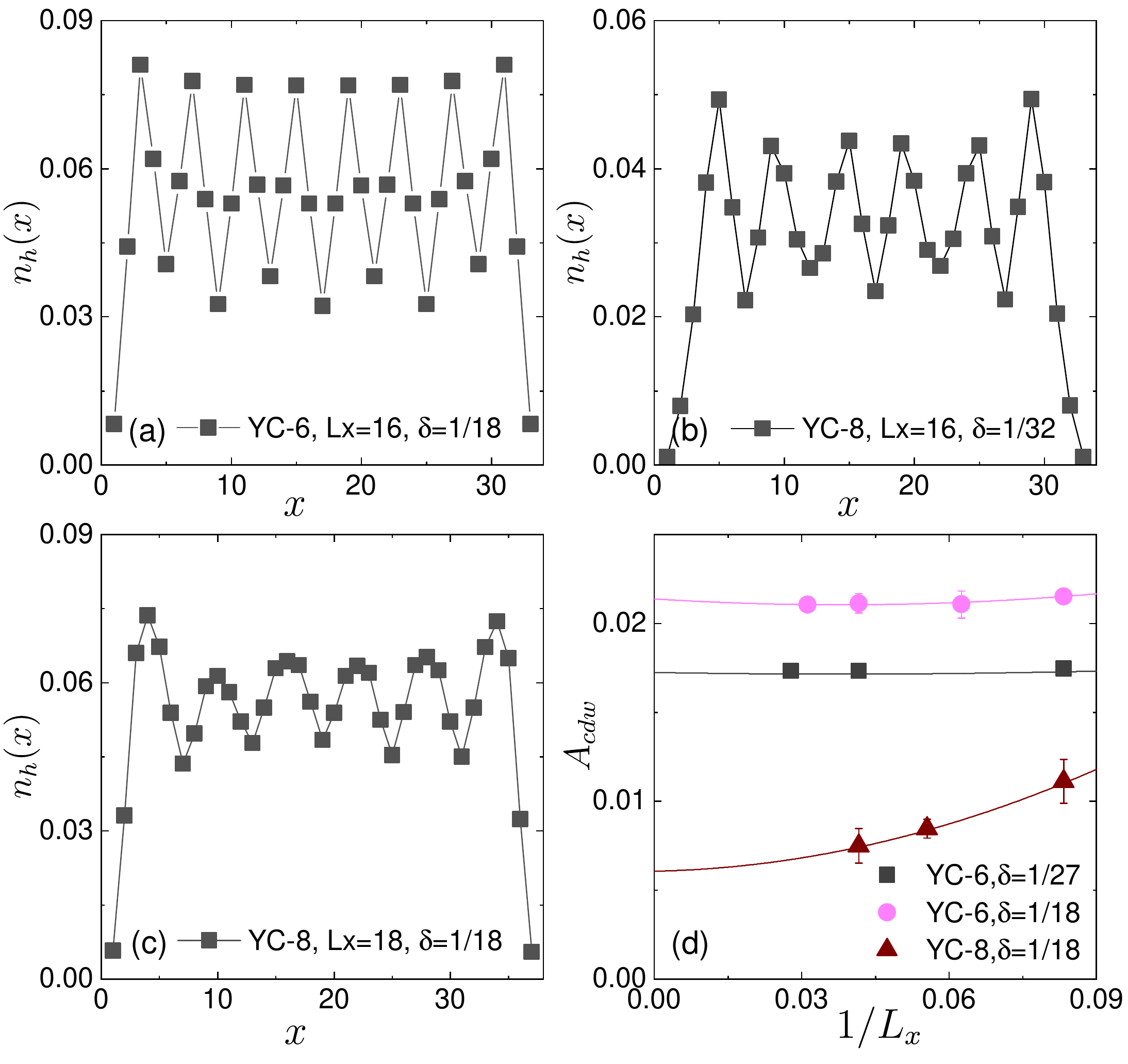}
\caption{(Color online) The charge density profile $n_h(x)$ (includes both A and B sites shown in Fig.\ref{Fig:Lattice}) for (a) YC-$6$ cylinder at hole doping concentration $\delta=1/18$, (b) YC-$8$ cylinder at $\delta=1/32$ and (c) YC-$8$ cylinder at $\delta=1/18$. (d) CDW order parameter $A_{cdw}$ at different hole doping concentration $\delta$ for both YC-$6$ and YC-$8$ cylinders.}\label{Fig:DCriticalSL_Acdw}
\end{figure}

\textit{Charge-density-wave order:} %
To describe the charge density properties of the ground state of the system, we consider the charge density profile $n_h(x,y)=1-n(x,y)$, where $n(x,y)$ is the electron density on site $i=(x,y)$. In Fig.\ref{Fig:DCriticalSL_CDW}, we show some examples of $n_{h}(x,y)$ at different doping concentrations $\delta$ for YC-$6$, YC-$8$ and YC-$10$ cylinders. Although the detailed pattern depends on both $\delta$ and the lattice geometry, a clear signature of CDW ordering is observed. This is similar with the the case of doping gapped spin liquid of AFM Heisenberg model on finite-width cylinders in Ref.\cite{Jiang2017}. Specifically, there is charge stripe (i.e., unidirectional CDW order) at low doping level, e.g., $\delta=1/18$ for YC-6 cylinders. For higher doping concentrations on YC-6 cylinders and lightly doped cases for both YC-8 and YC-10 cylinders, the CDW order is 2D which is similar to the one found in Ref.\cite{Jiang2017}.

Moreover, the doped system can be approximately divided into new larger emergent unit cells of either light-green stripes as shown in Fig.\ref{Fig:DCriticalSL_CDW}(a) or light-green raised spots in Fig.\ref{Fig:DCriticalSL_CDW}(b-f). Specially, the number of the emergent unit cells is equal to the number of doped holes for all doping concentrations, which demonstrates the absence of Cooper pairs in this lightly doped critical spin liquid on the Kagome lattice. Following the same procedure in Ref.\cite{Jiang2017}, we further calculate the averaged rung charge density defined as $n_h(x)=1/L_y\sum_{y=1}^{L_y}n_{h}(x,y)$ to determine whether there is a long-range CDW order in the thermodynamic limit. Fig.\ref{Fig:DCriticalSL_Acdw}(a-c) show some examples of $n_h(x)$ at various doping concentration for both YC-6 and YC-8 cylinders. Numerically, the existence of long-range CDW order of the lightly doped systems can be effectively determined by the amplitude, i.e., $A_{cdw}$, of averaged rung charge density modulation $n_h(x)$. For a given cylinder of length $L_x$, $A_{cdw}$ can be obtained by fitting the central-half region of $n_h(x)$using function $n_h(x)=n_0+A_{cdw}\cos(Qx+\theta)$, where $Q$ and $\theta$ are fitting parameters. Note that several data points close to both ends are removed in the fitting process in order to minimize the boundary effect. Fig.\ref{Fig:DCriticalSL_Acdw}(d) shows some examples of the finite-size extrapolations of $A_{cdw}$. It is clear that the presence of long-range CDW order is directly evidenced by the observed finite amplitude $A_{cdw}$ in the long-cylinder limit $L_x=\infty$. Similar with Ref.\cite{Jiang2017}, the ground state of lightly doped critical spin liquid can also be considered as a CDW of spinless holons with spin-charge separation, which is evidenced by the short-range spin-spin correlations and the absence of doping induced magnetic order.

\section{Doping chiral spin liquid}
It has long been proposed that doping a CSL or fractional quantum Hall state could give rise to topological superconductivity\cite{Laughlin1988,Wen1989}. This has been directly addressed by a recent DMRG study of the $t$-$J$ model supplemented by TRS breaking chiral interaction $J_\chi$ on the triangular lattice,\cite{Jiang2020CSL} which is defined by the Hamiltonian%
\begin{eqnarray}
    H= &-&t\sum_{\langle ij\rangle, \sigma}(\hat{c}^{\dagger}_{i\sigma}\hat{c}_{j\sigma}+h.c.) + J\sum_{\langle ij\rangle}(\bold{S}_i \cdot \bold{S}_j- \frac{\hat{n}_i \hat{n}_j)}{4} \nonumber \\
    &+&J_{\chi} \sum_{\bigtriangleup/\bigtriangledown} (\bold{S}_i\times \bold{S}_j)\cdot \bold{S}_k.
    \label{Eq:HamtJChi}
\end{eqnarray}
Large-scale DMRG simulations provide direct evidences that doping such a CSL on the triangular lattice can naturally yield topological superconductivity with $d\pm id$-wave pairing symmetry where the SC correlations are always dominant.

\begin{figure}
\centering
    \includegraphics[width=\linewidth]{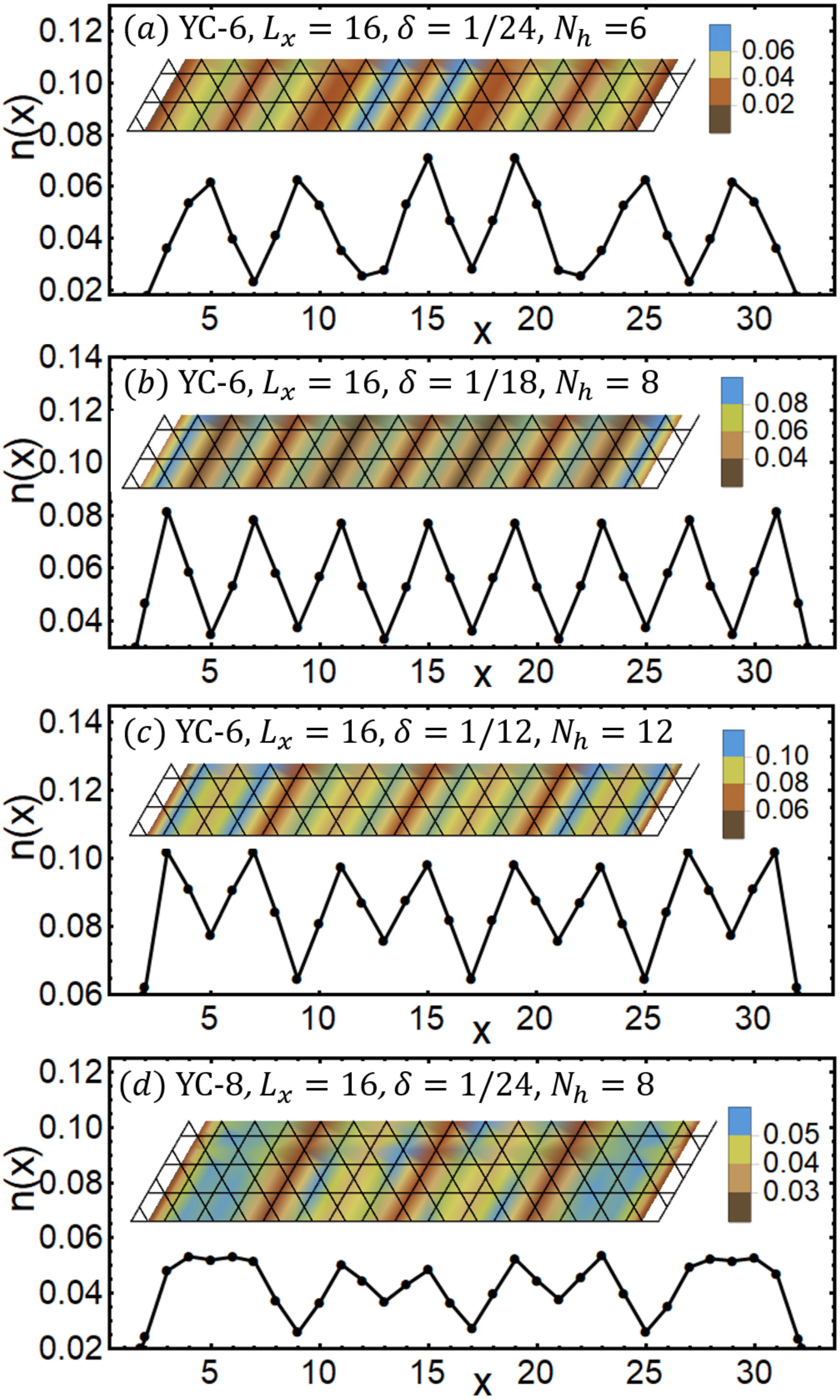}
\caption{(a-c) Hole density profile $n(x)$ for doped CSL on YC-6 cylinder at a serial of doping concentrations ranged from $\delta=1/24 \sim 1/12$. Inset is the density profile $n(x,y)$ of the same systems. (d) Hole density profile $n(x)$ on YC-8 cylinder at $\delta=1/24$ doping concentration.} \label{Fig:density_t_J_chi}
\end{figure}

Similarly, a number of recent DMRG studies have shown that CSL can also be realized on the Kagome lattice by either introducing both second and third-neighbor spin interactions \cite{Messio2012,He2014,Gong2014,Gong2015} or directly involving the TRS breaking chiral interaction $J_\chi$.\cite{He2015,Zhu2015,Bauer2014,Hu2015} Therefore, this raises a natural question that can topological superconductivity be realized in doping such a CSL on the Kagome lattice? In this paper, we directly address this question by studying the $t$-$J$-$J_\chi$ model defined in Eq.(\ref{Eq:HamtJChi}). Here $J_\chi$ denotes the spin SU(2) invariant spin scalar chirality living on both up and down triangles of the Kagome lattice where sites $i$, $j$ and $k$ in each triangle following the clockwise order. The local Hilbert space is constrained by the no-double occupancy condition, $n_i \le 1$. At half-filling, Eq.(\ref{Eq:HamtJChi}) reduces to the $J$-$J_\chi$ model, which has a stable gapped CSL in a larger parameter region $ 0.05\pi \lesssim $ arctan$| J_\chi/J | \lesssim \pi/2$, i.e., 0.158 $\leq |J_\chi/J|$.\cite{Bauer2014} For the present study, we focus on the lightly-doped case with a small hole doping concentration $\delta$. We set $J=1$ as an energy unit and consider $t=3$ and $J_\chi=0.5$, which is deep in the CSL phase of the $J$-$J_\chi$ model. We use complex-valued DMRG with spin SU(2) symmetry to simulate the $t$-$J$-$J_\chi$ model in  Eq.(\ref{Eq:HamtJChi}). We keep up to $m=5000$ SU(2) states (effectively $15000$ U(1) states) for YC-6 cylinders with at least 30 sweeps to obtain accurate results with typical truncation error $\epsilon\leq 10^{-6}$. For YC-8 cylinders, $m=7000$ SU(2) ($\sim 21000$ U(1)) states are kept to obtain reliable results.

To determine the ground state properties of the system upon light doping, we first calculate the charge density profile $n_h(x,y)$, where four examples are shown in the insets of Fig.\ref{Fig:density_t_J_chi} for YC-6 and YC-8 cylinders. It is clear that the doped holes form unidirectional CDW ordering, i.e., charge stripes. Similar with previous sections, the system of lightly doped CSL can also be approximately divided into new larger emergent unit cells. While it is not very straightforward to establish the absence of hole pairs for $1/12$ doping concentration in Fig.\ref{Fig:density_t_J_chi}(c), it is clear that for YC-6 cylinders with $\delta=1/24$ and $1/18$ doping concentration as shown in Fig.\ref{Fig:density_t_J_chi}(a-b), the number of emergent unit cells is equal to the number of doped holes. Similar charge stripes with one doped hole are observed in the bulk of the YC-8 cylinder with $\delta=1/24$ doping concentration (Fig.\ref{Fig:density_t_J_chi}(d)), indicating that the charge stripes emerged from the doped CSL is robust as the systems become wider. 
Moreover, there is no doping induced magnetic order, from which we could conclude that the ground state of the system forms a stripe crystal of spinless holons with spin-charge separation. This is very similar with the stripe crystal observed in lightly doped $t$-$J$ model on YC-6 cylinders with doping concentrations $\delta\leq 1/18$ as shown in Fig.\ref{Fig:tJ_Ni}. To further determine whether the CDW order is long-ranged, we also calculate the averaged rung charge density $n_h(x)$, where examples are shown in Fig.\ref{Fig:density_t_J_chi}. It is clear that $n_h(x)$ has a strong spatial oscillation whose amplitude barely decays or even increases from the boundary to the bulk of the system. Therefore, while we cannot perform the finite-size extrapolation, it is very likely that such a stripe CDW order appears to be long-ranged in the $L_x=\infty$ limit for YC-6 and YC-8 cylinders. Nevertheless, whether the stripe crystal still survives for higher doping concentrations on YC-8 cylinders or wider systems, or whether it will be replaced by the 2D Wigner crystal or other phases, needs to be determined in a further study.

\begin{figure}
\centering
    \includegraphics[width=1\linewidth]{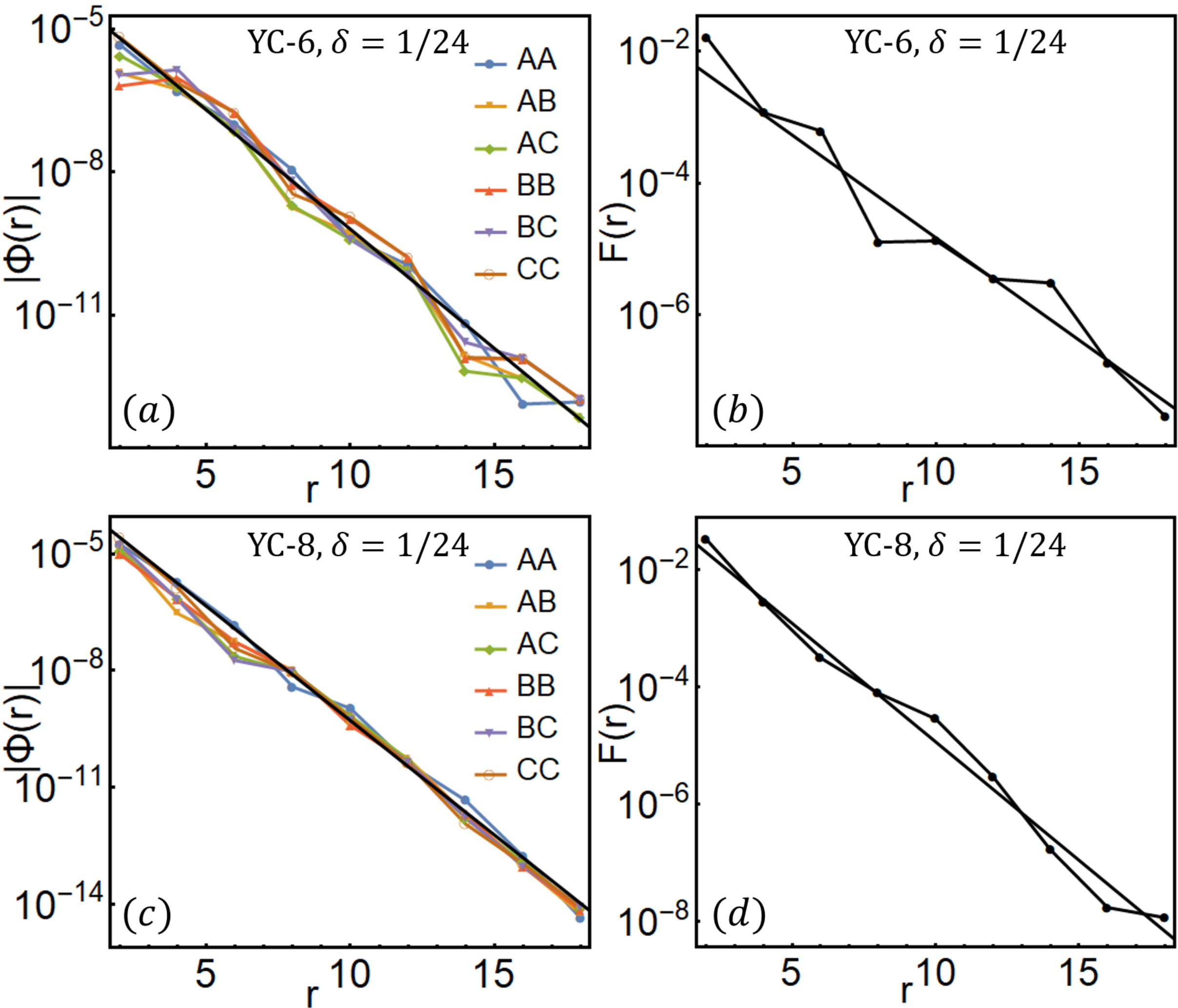}
\caption{(a-b) The pair field correlation function $\Phi(r)$ and spin-spin correlation function $F(r)$ for lightly doped CSL on YC-6 cylinder with $\delta=1/24$ doping concentration. (c-d) The correlation functions $\Phi(r)$ and $F(r)$ on YC-8 cylinder with doping $\delta=1/24$. }\label{Fig:cor_t_J_chi}
\end{figure}

As there is no indication of Cooper pairs in the ground state of the lightly-doped system, it is natural to expect the absence of superconductivity. To demonstrate this, we have calculated the SC pair-field correlations $\Phi(r)$ where examples for YC-6 and YC-8 cylinders at $\delta=1/24$ doping concentration is shown in Fig.\ref{Fig:cor_t_J_chi}(a) and (c). It is clear that all $\Phi_{\alpha\beta}(r)$ decay exponentially at long-distances and can be well fitted by the exponential function $\Phi(r)\sim e^{-r/\xi_{sc}}$. For both YC-6 and YC-8 cases, the extracted SC correlation lengths are $\xi_{sc}\sim 0.8$ unit cell, nearly independent on bond types $\alpha\beta$. The spin-spin correlation functions $F(r)$ are also short-ranged in the lightly doped region as shown in Fig.\ref{Fig:cor_t_J_chi}(b) and (d), which can be fitted by the exponential function $F(r)\sim e^{-r/\xi_s}$ with a short correlation length $\xi_s\sim 1.3$ and $\xi_s\sim 1.1$ unit cell for YC-6 and YC-8 cases, respectively. Similar behaviors are also observed on YC-6 cylinder at $\delta=1/18$ doping concentration.

\section{Summary and conclusion}
We have reviewed and studied the consequence of lightly-doped QSLs on the Kagome lattice. In the context of $t$-$J$ model and its generalizations, we show that doping distinct types of QSLs on the Kagome lattice all give rise to insulating phase with long-range CDW order. Although the pattern of CDW ordering, i.e., stripe crystal or Wigner crystal, depends on both the lattice geometry and hole doping concentrations, it is the crystal of spinless holons with spin-charge separation. While the strong resemblance of doping gapped and critical spin liquids suggests that the gapless spin liquid can also be stabilized in the Kagome Heisenberg antiferromagnet in two dimensions, we suspect that the observed CDW ordering might be intrinsic to the geometry of Kagome lattice, given the fact the similar state has also been observed in doping the qualitatively distinct chiral spin liquid. This is further supported by our preliminary results of doping magnetically ordered state on the Kagome lattice, which gives rise to similar insulating CDW phase.

In light of our numerical observations, it is worth asking what is the reason that the holons crystallize, rather than forming the superconducting or other metallic states. It is clear from previous studies that the QSL and a number of possible valence-bond-crystalline phases on the kagome lattice are nearly degenerate. It has also been shown in Ref.\cite{Jiang2017} that a strong and extended pattern of valence-bond-crystalline order can be induced by doping two holes into moderate length Kagome cylinders, it is hence natural to infer that the holon effective mass would be strongly renormalized and increased, and effective repulsive interactions between holons implied by the doping induced valence-bond-crystal order could naturally lead to crystallization.\cite{Jiang2017} Based on our numerical observations, it seems natural to expect that the corollary of Ref.\cite{Jiang2017} could also apply to doping critical and chiral spin liquids on the Kagome lattice.

It is worth mentioning that our numerical results have shown that the SC correlations can be significantly enhanced by introducing the next-nearest-neighbor electron hopping $t^\prime$. It is thus natural to expect that the introduction of even longer-range electron hopping terms could further increase the SC correlations, where true long-range superconductivity might be eventually realized. This can be considered as a potential pathway to realize superconductivity or other metallic states in doping Kagome materials, including herbertsmithite ZnCu$_3$(OH)$_6$Cl$_2$ and Zn-substituted barlowite Cu$_3$Zn(OH)$_6$FBr,\cite{Han2012,Fu2015,Liu2015,Hering2017,Feng2017,Smaha2020}, by applying strain or pressure.

{\it Acknowledgments: } We would like to sincerely thank Steve Kivelson for insightful discussions, invaluable suggestions and generous help to improve the manuscript. We are grateful to Thomas Devereaux and Steve Kivelson for collaborations that deepened our interest and understanding of the subject. This work was supported by the Department of Energy, Office of Science, Basic Energy Sciences, Materials Sciences and Engineering Division, under Contract DE-AC02-76SF00515. Parts of the computing for this project was performed on the National Energy Research Scientific Computing Center (NERSC), a US Department of Energy Office of Science User Facility operated under Contract No.DE-AC02-05CH11231. Parts of the computing for this project was performed on the Sherlock cluster. The calculations in Sec.III are performed using the high-performance matrix product state algorithm library GraceQ/MPS2\cite{GraceQ}

\bibliography{Refs}

\end{document}